# Beyond 100 GHz Silicon Mach-Zehnder Modulators Enabled by Spatially Selective Optimization

XINGYU LIU,[1,†] FENGXIN YU,[2,†] XIAO HU,[2,*] ZIXUAN CAI,[1] HAIBIN ZHAO,[1] HAIWEN CAI,[2,3] AND WEI CHU[2,4]

[1]*College of Future Information Technology, Fudan University, Shanghai 200433, China*
[2]*Zhangjiang Laboratory, Shanghai 201210, China*
[3]*caihw@zjlab.ac.cn*
[4]*chuwei@zjlab.ac.cn*
[†]*These authors contributed equally.*
**hxhustoe@163.com*

**Abstract:** The growing demand for higher-capacity optical interconnects in artificial intelligence and high-performance computing calls for silicon modulators with electro-optic bandwidths beyond 100 GHz. Optical and electrical functions in silicon phase shifters respond differently to local design changes, creating an opportunity to optimize the conductive access region while preserving the central phase-shifting geometry. Here, we demonstrate a stepped-slab junction-access structure on a 300 mm silicon photonics platform. By increasing the conductive cross-section away from the optically sensitive phase-shifting region, the structure reduces junction-access resistance. The 2 mm-Mach–Zehnder modulator achieves a directly measured 3 dB electro-optic bandwidth of 104 GHz and a half-wave voltage of 7.5 V. It also exhibits open eye diagrams at 170 GBaud eight-level pulse-amplitude modulation (PAM8), corresponding to 510 Gbit/s. This process-compatible spatially selective strategy provides a practical route towards silicon modulators for next-generation 400 Gbit/s-per-lane optical interconnects.

## 1. Introduction

The continued growth of artificial intelligence (AI) and high-performance computing (HPC) is driving optical interconnects towards increasingly higher per-lane capacities. In response to this trend, the IEEE 802.3 Working Group chartered a study effort in March 2026 to examine 400 Gb/s-per-lane electrical interconnects and single-mode-fiber optical interconnects with reaches up to 500 m, highlighting the growing demand for high-capacity links in next-generation computing systems [1]. At the 400 Gb/s-per-lane level, higher-order intensity-modulation formats such as four-level pulse-amplitude modulation (PAM4) require a symbol rate of approximately 200 GBaud, corresponding to a Nyquist frequency near 100 GHz and consequently imposing stringent bandwidth requirements on optical transmitters [1-3].

Silicon photonics provides a practical platform for this scaling because of its complementary metal–oxide–semiconductor (CMOS) compatibility, wafer-scale manufacturability and high integration density [2-4]. Among silicon photonic modulator architectures, Mach–Zehnder modulators (MZMs) are among the most mature and widely adopted solutions for high-speed optical interconnects, benefiting from their broadband, non-resonant operation, relatively low sensitivity to wavelength variations, and well-established travelling-wave and push–pull configurations. Recent roadmapping studies indicate that travelling-wave MZMs currently dominate the commercial silicon-photonics modulator market, while MZMs have already been adopted in data-centre intensity-modulation direct-detection transceivers [3]. At the same time, the relatively weak plasma-dispersion effect in silicon requires conventional MZMs to employ millimetre-scale phase shifters to accumulate sufficient phase shift. This long interaction length provides practical modulation efficiency but also increases the accumulated junction capacitance and microwave attenuation, constraining further bandwidth scaling.

Considerable efforts have therefore explored different routes to extend the bandwidth of silicon modulators. One approach relies on directly engineering the optical interaction, by employing slow-light enhancement or resonant structures to increase the effective modulation strength and shorten the required phase-shifter length. Silicon slow-light MZMs with sub-millimetre phase shifters have demonstrated electro-optic (EO) bandwidths exceeding 110 GHz [5] and 400 Gbps/λ operation [6], while more compact resonant architectures, including microring and photonic-crystal nanocavity modulators, can reduce the active length to tens of micrometres [7-9]. These approaches substantially relax the electrical loading associated with long phase shifters, but introduce additional considerations related to fabrication variations, wavelength control, thermal stability and optical passband. Another approach preserves the conventional MZM architecture while instead shaping its high-frequency response. On-chip inductive peaking [10,11], impedance peaking [12] and RC equalization [13] can reshape the EO frequency response, while optical-domain equalization based on crossing waveguides or counter-doping can compensate frequency-dependent modulation response by reversing the modulation direction [14-17]. Some equalization schemes, however, improve bandwidth flatness partly by suppressing the low-frequency response and thus reducing the effective modulation response. At the transmitter level, driver–MZM co-design provides a complementary system-level strategy, in which the driver, inductive network, termination impedance and modulator are jointly optimized to balance high-speed performance and energy efficiency [18]. Within this transmitter-level co-design framework, a P–N dual-drive scheme with independently engineered P/N frequency responses has enabled an all-silicon transmitter to operate at 200 GBaud with 2.5 pJ/bit [19]. More recently, a silicon MZM driven by a commercial high-speed SiGe driver has demonstrated 400 Gbps/lane PAM4 transmission [20], establishing the practical feasibility of this mature modulator architecture for the emerging 400 Gbps/lane regime.

Taken together, these advances show that silicon MZMs can operate in 100 GHz-class and 400 Gbps/lane systems through optical-structure engineering or electrical–photonic co-design. However, such approaches may introduce additional architectural, wavelength-control or co-optimization constraints, while some equalization schemes reshape rather than intrinsically extend the modulator response. A wider intrinsic bandwidth also reduces the degree of frequency compensation required from surrounding electrical and optoelectronic components. This motivates a device-level strategy that revisits the internal structure of the conventional MZM to improve its intrinsic high-frequency performance. In particular, the series resistance of the p–n junction access region offers a local electrical design degree of freedom distinct from external circuit-level compensation. Optical and electrical functions in a silicon phase shifter exhibit different spatial sensitivities to local design changes. The optically active region must maintain adequate carrier–optical-mode overlap while suppressing free-carrier absorption, whereas the conductive access region is governed mainly by its conductive cross-section and material resistivity. This distinction allows the access region to be selectively optimized while minimizing perturbation to the optically sensitive core. Such spatially selective optimization offers a route to improving intrinsic bandwidth without fundamentally altering the MZM operating principle.

In this work, we experimentally demonstrate this concept in a 2 mm O-band silicon MZM fabricated on a 300 mm silicon photonics platform using a stepped-slab junction-access structure. By increasing the conductive cross-section in the junction-access region while preserving the active phase-shifting geometry, the structure reduces access resistance and the associated RF attenuation and RC loading. Without requiring heterogeneous electro-optic materials, ultrashort resonant phase shifters or additional fabrication modules, the device achieves EO bandwidths of 94 GHz at −3 V and 104 GHz at −5 V, with a modulation efficiency of $V_\pi \cdot L$ = 1.5 V·cm ($V_\pi$ = 7.5 V) and an on-chip insertion loss of 4.5 dB. The device also exhibits open eye diagrams at 170 GBaud PAM8, corresponding to 510 Gbit/s. This work establishes a process-compatible approach to extending the bandwidth of conventional silicon MZMs into the 100 GHz regime, offering increased bandwidth margin for next-generation 400 Gbit/s-per-lane silicon optical transmitters.

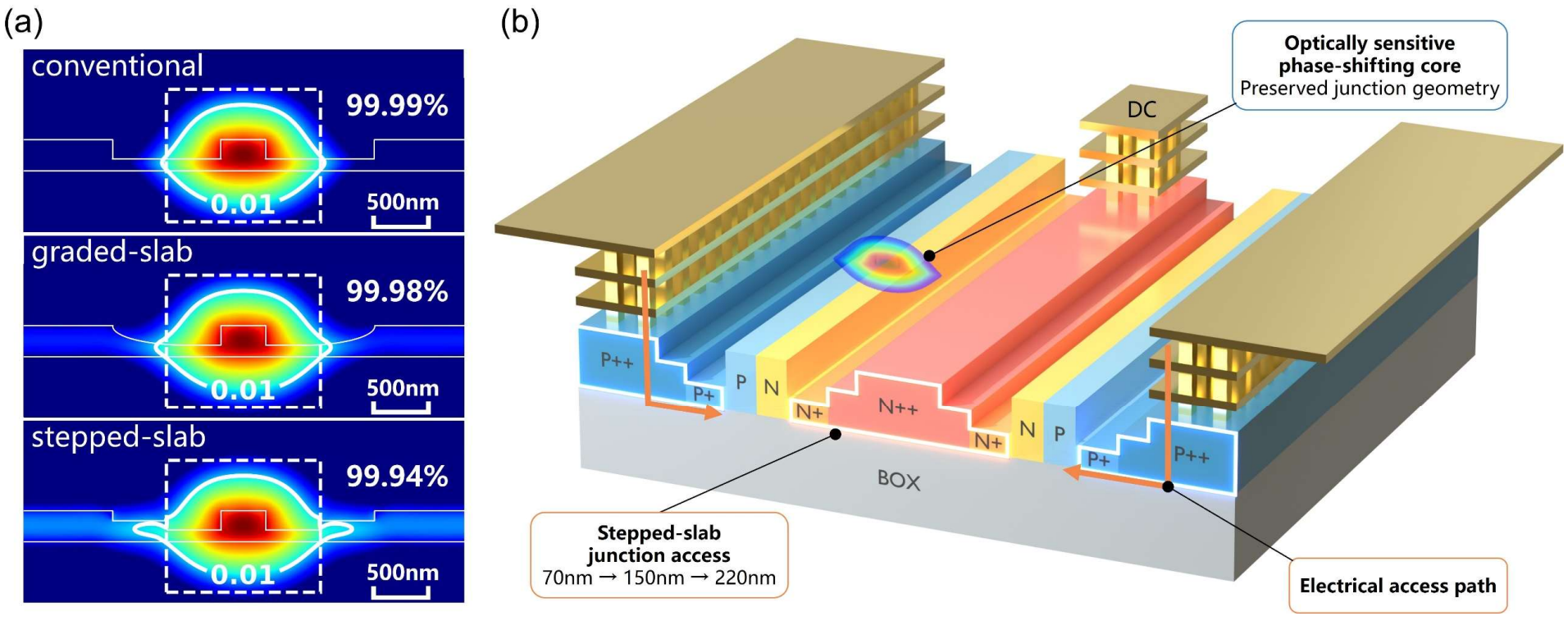

Fig. 1. Modulator design. (a) Optical field at different ridge cross-sections. (b) Three-dimensional schematic diagram of the designed modulator.

## 2. Spatially selective optimization

The distinct spatial sensitivities of optical and electrical functions provide a design degree of freedom for improving conventional silicon MZMs. As shown in Fig. 1(a), a carrier-depletion phase shifter based on a conventional rib waveguide comprises a central rib and two lateral slab regions extending towards the electrical contacts. Both regions contribute to the current path and hence the p–n junction series resistance, whereas the optical mode is concentrated in the rib and decays towards the outer slabs. We quantify this confinement using the fraction of optical power within the central waveguide region:

$$\Gamma_{core} = \frac{\iint_{A_{core}} S_Z(x,y)\, dA}{\iint_{A_{section}} S_Z(x,y)\, dA} \tag{1}$$

where S_z is the longitudinal component of the optical Poynting vector. As indicated by the white dashed box in Fig. 1(a), 99.99% of the optical power is confined to the central rib and its immediately adjacent region. This spatial distribution suggests that the electrical access region can be thickened progressively away from the optical core, increasing its conductive cross-section where optical-field overlap is weak. An idealized continuously graded slab retains 99.98% of the optical power within the same central region. This profile can be approximated by a fabrication-compatible stepped-slab geometry (Fig. 1(b)), in which the silicon thickness increases from a 70 nm inner slab adjacent to the rib to a 150 nm intermediate slab and then to the full 220 nm device-layer thickness towards the electrical contacts. Beyond this geometric configuration, three levels of PN junction doping (light, moderate, and heavy) are used to balance optical and electrical performance.

To examine how the location of electrical optimization affects device performance, we compare two pathways from the same baseline geometry: increasing the carrier concentration in the central active region and modifying the outer slab geometry using a stepped profile. Both designs share a 380 nm-wide waveguide, with the electrical contacts positioned 1000 nm from the central waveguide edge. Carrier distributions derived from Technology Computer Aided Design (TCAD) simulations are used consistently for small-signal electrical extraction and finite-difference eigenmode (FDE) analysis, enabling a comparison of optical loss, junction resistance, capacitance and modulation efficiency.

Figs. 2(a) and 2(d) reveal distinct optical constraints for the two pathways. Increasing the carrier concentration enhances free-carrier absorption, whereas reducing the trench distance—the lateral distance between the waveguide edge and the stepped-slab transition—strengthens evanescent-field coupling to the outer slab. The propagation loss increases markedly when the trench distance falls below approximately 500 nm. Thus, both approaches have an optical-loss constraint, but the underlying mechanisms differ: absorption in the doped active region versus optical-field leakage towards the modified access region.

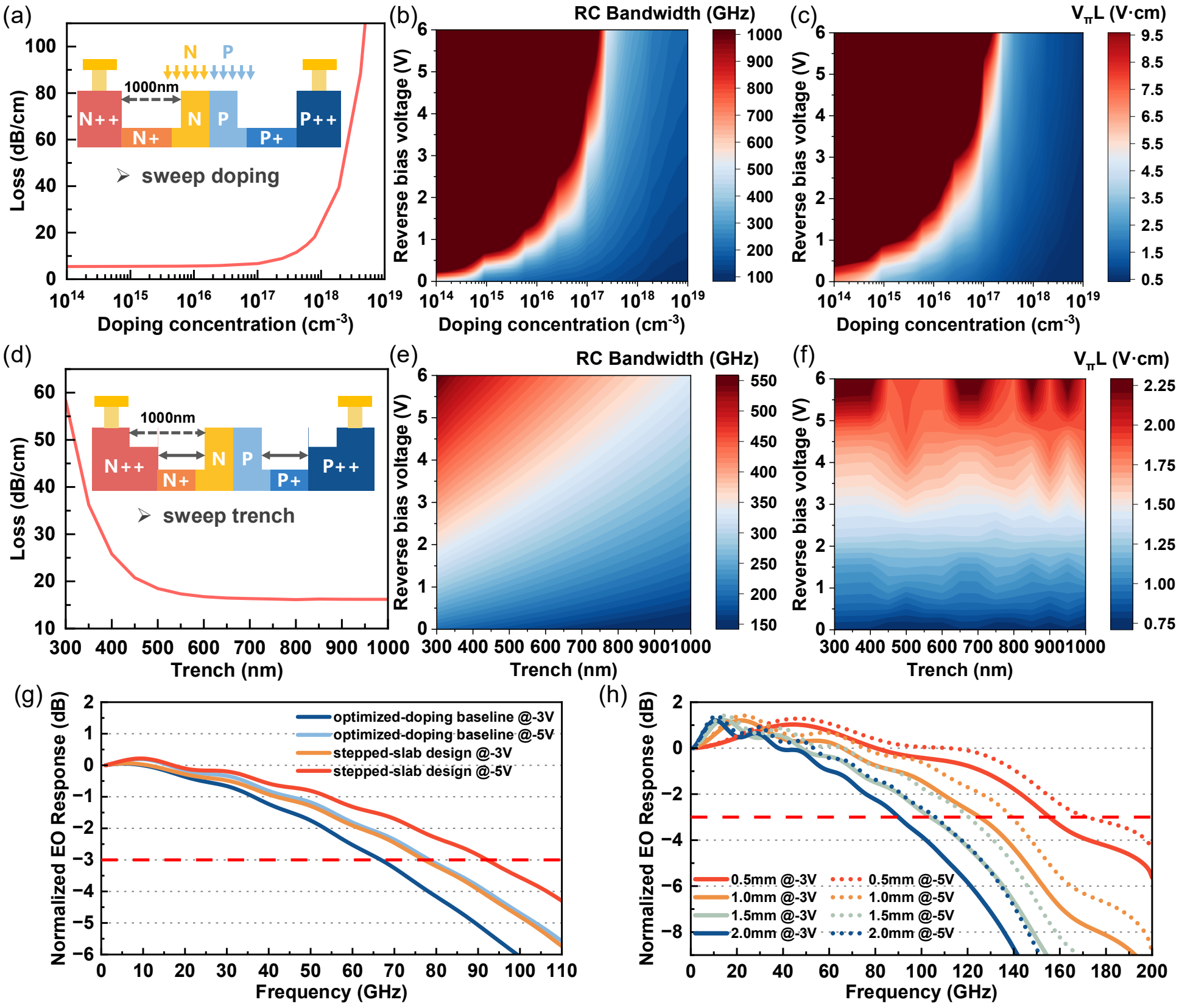

Fig. 2. Design parameter sweeps and simulation results. (a), (b), (c) Optical loss, RC bandwidth, and Vπ·L vs. doping concentration and reverse bias. (d), (e), (f) Optical loss, RC bandwidth, and Vπ·L vs. trench distance and reverse bias. (g) EO response of 2 mm MZM at various bias voltages. (h) EO response of MZMs with various lengths under termination mismatch.

The electrical and modulation results further distinguish the two approaches, as shown by the intrinsic PN-junction RC bandwidth (Figs. 2(b) and 2(e)) and modulation efficiency (Figs. 2(c) and 2(f)). The series resistance RS and junction capacitance CPN, extracted as detailed in Supplementary Information S1, determine the intrinsic RC bandwidth, $f_{RC} = 1/2\pi R_S C_{PN}$. Increasing the active-region carrier concentration reduces both the RC bandwidth and Vπ·L, indicating a trade-off between intrinsic electrical speed and modulation efficiency. At low doping, the RC bandwidth exceeds 1000 GHz, but Vπ·L remains above 9 V·cm, far exceeding the typical 1–2 V·cm range of silicon MZMs. Higher doping improves modulation efficiency but reduces the RC bandwidth and increases free-carrier absorption (Fig. 2(a)). By contrast, reducing the trench distance brings the thicker, more conductive slab closer to the waveguide, lowering $R_S$ and thereby increasing the intrinsic RC bandwidth, while CPN and Vπ·L remain nearly unchanged because the stepped transition is separated from the central depletion region. For the geometry sweep, the active-region carrier concentration is fixed at $6.7\times10^{17}$cm$^{-3}$, corresponding to Vπ·L ≈ 1.5V·cm at −3 V.

We next assess how these electrical differences translate into the EO response using a p–n-junction-loaded transmission-line model [21,22], with the microwave electrode characteristics obtained from ANSYS HFSS (Supplementary Information S2). The 2 mm MZM adopts a series push–pull configuration to reduce junction capacitance and T-shaped slow-wave electrodes to match the microwave and optical group velocities. Fig. 2(g) compares the normalized EO responses of the stepped-slab device with a 500 nm trench and the uniform-slab reference, with the same optimized doping profile in both devices. With a matched termination, the reference MZM exhibits EO bandwidths of 67 GHz at −3 V and 79 GHz at −5 V, whereas the stepped-slab design reaches 78 GHz and 92 GHz, respectively. These gains arise from reduced electrical loading associated with the lower series resistance, which both mitigates dissipative loss and reduces the voltage dropped across the junction resistance at high frequencies, leaving a larger fraction of the drive voltage across the depletion capacitance (Supplementary Information S2).

We further examine termination-mismatch-induced peaking as a distinct mechanism that can increase the apparent 3 dB EO bandwidth. As shown in Fig. 2(h), a 50 Ω traveling-wave electrode with a 40 Ω termination produces approximately 1 dB of peaking, increasing the 2 mm device normalized EO bandwidth from 78 to 90 GHz at −3 V and from 92 to 106 GHz at −5 V. At −3 V, the corresponding bandwidths reach 104, 126, and 156 GHz for 1.5 mm, 1 mm, and 0.5 mm devices, respectively, and approach 200 GHz for even shorter devices (Supplementary Information S3). The response peak shifts from approximately 11 GHz for a 2 mm electrode to 45 GHz for a 0.5 mm electrode, consistent with reflected-microwave phase accumulation and the approximate scaling $f_{peak} \propto 1/L$ (Supplementary Information S3). This mismatch-induced enhancement should be distinguished from the bandwidth improvement achieved by reducing the junction-access resistance.

Together, these simulations establish the design rationale for the stepped-slab MZM: the optical mode distribution identifies where additional conductive silicon can be introduced, while the electrical and EO analyses quantify the resulting bandwidth benefit and optical-loss constraint. The 500 nm trench stepped-slab design is therefore selected for device fabrication and experimental validation.

## 3. DC and Small-Signal Characterization of the Stepped-Slab MZM

To experimentally validate the proposed spatially selective junction-access design, 2 mm MZMs were fabricated on a 300 mm silicon photonics platform using the process-compatible stepped-slab structure described above. Fig. 3(a) shows the fabricated 300 mm wafer and a photograph of a representative device. The fabricated devices were characterized at both the individual-device and wafer levels.

Fig. 3(b) compares the transmission spectra of the grating-coupler (GC) reference and the MZM across the O-band. After accounting for the fibre-link and GC contributions, the separation between their spectral envelopes near 1310 nm yields an on-chip insertion loss of 4.5 dB. The spectra remain broad across the relevant O-band operating window, without a pronounced wavelength-dependent loss feature. The modulation efficiency was extracted from the bias-dependent spectral shift of the MZM shown in Fig. 3(c). The unequal-arm interferometer has a free spectral range (FSR) of approximately 3.8 nm. A common bias of −3 V was applied, with differential voltages Vd = 0–1 V distributed symmetrically between the two arms as −3 V ± Vd/2; fitting the odd spectral shift normalized by the FSR yields Vπ = 7.5 V and Vπ·L = 1.5 V·cm. The EO response was measured with a Keysight N4372E and de-embedded to the calibrated radio-frequency (RF) reference plane. Fig. 3(d) shows the measured

EO responses under bias voltages ranging from −1 to −5 V. The 3 dB EO bandwidth reaches 94 GHz at −3 V and increases to 104 GHz at −5 V. The measured bandwidth agrees well with the simulations using the distributed EO model in Figs. 2(g) and 2(h), confirming the effectiveness of the reduced access resistance enabled by the stepped-slab structure in improving the intrinsic bandwidth of the MZM. Additional comparisons between measurements and simulations can be found in Supplementary Information S4.

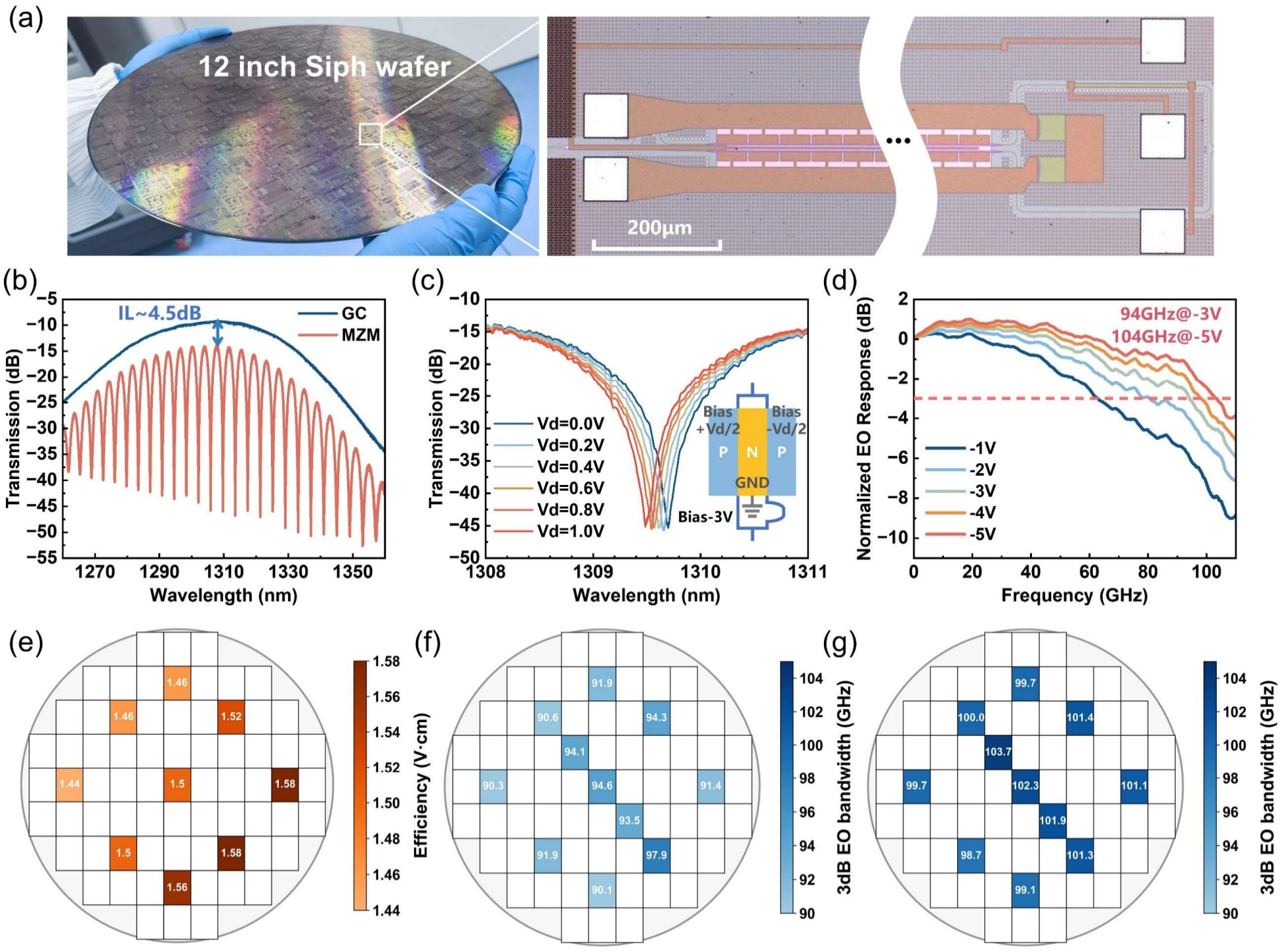

Fig. 3. Device fabrication and static characterization. (a) Photograph of a 12 inch wafer and microscope photograph of the fabricated modulator. (b) Measured optical transmission of GC-GC and MZM. (c) Phase shifter spectra. (d) Normalized EO responses from −1 V to −5 V. (e) Wafer map of Vπ·L. (f), (g) Wafer map of 3 dB EO bandwidth at −3 V and −5 V bias.

Wafer-level measurements were used to assess device uniformity across the 300 mm wafer. As shown in Fig. 3(e), the mean Vπ·L is 1.5 V·cm, with a standard deviation of 0.053 V·cm. Fig. 3(f) and 3(g) show the wafer-level distributions of the 3 dB EO bandwidth at −3 V and −5 V, respectively. The mean bandwidths are 92.8 and 100.8 GHz, with standard deviations of 2.4 and 1.6 GHz, respectively. These results indicate relatively low within-wafer variation in both modulation efficiency and EO bandwidth, supporting the suitability of the stepped-slab structure for implementation on a 300 mm silicon photonics platform.

## 4. High-Speed Performance Characterization

Large-signal modulation was characterized using the setup shown in Fig. 4(a). A Keysight M8199B 256 GSa/s arbitrary waveform generator directly drove the MZM through a 110 GHz single-ended RF probe without an additional RF driver. The modulator was biased at −2.5 V during the large-signal measurements. Each grating coupler introduced approximately 5 dB of insertion loss, and the optical output was amplified by a praseodymium-doped fiber amplifier (PDFA), providing approximately 4 dBm at the digital communication analyzer (DCA) input.

Fig. 4(b) presents the measured eye diagrams at different modulation formats and symbol rates. At 112 GBaud, PAM4 operation corresponds to 224 Gbit/s and yields an instrument-reported TDECQ of 0.85 dB, an outer extinction ratio of 1.16 dB and an RLM of 0.96. The relatively modest extinction ratio is primarily associated with the limited high-frequency drive swing and the single-ended measurement configuration [20,23,24]. In particular, the nominal 2.7-Vpp AWG output is estimated to decrease to approximately 1.3 Vpp at 128 GBaud [25] after accounting for the electrical response of the measurement path. Open eye diagrams were also obtained at 180 GBaud PAM4, 180 GBaud PAM6 and 170 GBaud PAM8, corresponding to 360 Gbit/s, an equivalent rate of 465 Gbit/s, and 510 Gbit/s, respectively. A 15 tap feed-forward equalizer (FFE) was used at 112 GBaud, whereas a 30 tap FFE was employed for symbol rates of 170 GBaud and above. In addition, Fig. 4(c) shows the measured bit error rates (BERs) of PAM4 signals at different data rates. At 300 Gbit/s (150 GBaud), the BER is below the hard-decision forward error correction (HD-FEC) threshold of $3.8\times10^{-3}$. At 360 Gbit/s (180 GBaud), it remains below the soft-decision FEC (SD-FEC) threshold of $4.5\times10^{-2}$.

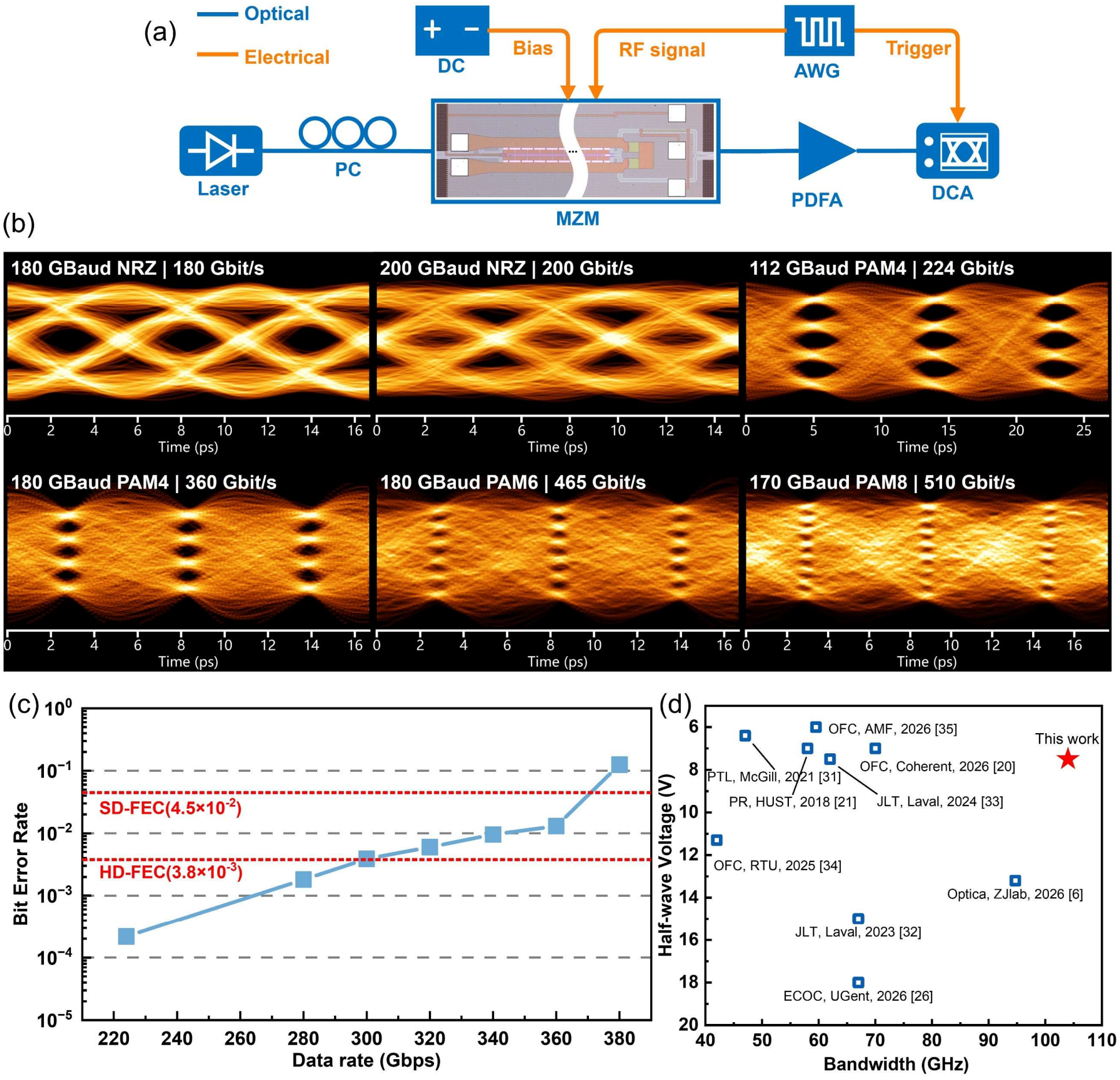

Fig. 4. High-speed operation characterization and performance comparison. (a) Experimental setup for high-speed characterization. (b) Eye diagrams at various modulation formats and data rates. (c) BER versus PAM4 signal rate. (d) Review of silicon MZMs: EO bandwidths and Half-wave Voltage.

## 5. Discussion

In this work, we demonstrate a 2 mm silicon MZM with a directly measured 3 dB EO bandwidth of 104 GHz, which, to the best of our knowledge, represents the highest directly measured 3 dB EO bandwidth reported for a conventional silicon MZM without relying on resonant enhancement or integrated equalization to extend its bandwidth. This improvement is enabled by a spatially selective design strategy that exploits the different spatial sensitivities of optical modulation and electrical conduction: instead of further modifying the optically sensitive junction core, the conductive slab access region is selectively engineered to reduce the junction series resistance. Table 1 summarizes the performance parameters of recent representative silicon-based MZMs, including devices reporting EO bandwidths above 110 GHz or single-lane data rates of 400 Gbit/s. The present device combines 104 GHz EO bandwidth with a relatively low Vπ of 7.5 V, a Vπ·L of 1.5 V·cm, and open eye diagrams at 170 GBaud PAM8 (510 Gbit/s), while retaining the conventional 2 mm MZM architecture. As summarized in Fig. 4(d), these characteristics place the present device in the high-bandwidth, low-Vπ region and provide a competitive combination of bandwidth, drive efficiency and demonstrated data rate.

**Table 1. Comparison of Key Performance Metrics for Si-based MZMs.**

| Ref | Op.band | Key structural feature | Length (mm) | IL (dB) | VπL (V·cm) | Vπ (V) | 3 dB BW (GHz) | Speed (Gbps) |
|---|---|---|---|---|---|---|---|---|
| 2023 [5] | Cband | Photonic crystal | 0.124 | 5.4 | 0.96 | 77.4 | 110 (@-3V) | 112 (NRZ) |
| 2023 [32] | Cband | Segmentation | 2 | 2 | 3 | 15 | 67 (@-4V) | 360 (PAM8) |
| 2024 [33] | Cband | Micro-capacitors | 4 | 4.5 | 3 | 7.5 | 61 (@-4V) | 240 (PAM4) |
| 2025 [16] | Cband | Frequency equalization | 0.9 | 4.3 | 4.86 | 54 | 110 (@-6V) | 140 (NRZ) |
| 2025 [17] | Oband | Frequency equalization | 2.58 | 2.1 | 6.48 | 25.1 | 90 | 200 (PAM4) |
| 2025 [34] | Cband | - | 1.5 | 3.6 | 1.7 | 11.3 | 42 (@-2.5V) | 375 (PAM6) |
| 2026 [6] | Oband | Photonic crystal | 0.5 | 2.4 | 0.66 | 13.2 | 94.7 | 405 (PAM8) |
| 2026 [35] | Oband | Doping engineering | 2.25 | 2.45 | 1.36 | 6 | 59.5 | 336 (PAM8) |
| 2026 [20] | - | - | - | - | - | ≈7 | >67 | 420 (PAM4) |
| 2026 [26] | Cband | - | 1 | 1.5 | 1.8 | 18 | >67 | 425 (PAM4) |
| **This work** | **Oband** | **Stepped-slab** | **2** | **4.5 (@0V)** | **1.5 (@-3V)** | **7.5** | **94 (@-3V)**<br>**104 (@-5V)** | **360 (PAM4)**<br>**465 (PAM6)**<br>**510 (PAM8)** |

* IL: Insertion Loss; BW: Bandwidth; -: Data is not available.

Despite these advantages, the bandwidth enhancement comes at the cost of an insertion loss of approximately 4.5 dB, which is slightly higher than that of other MZMs with comparable lengths in Table 1, mainly associated with the stepped-slab geometry. This penalty requires higher launched optical power to maintain the same received power, increasing the laser power budget. Potential mitigation through high-power O-band laser sources and high-sensitivity photodetectors, and reduced coupling losses at optical interfaces and propagation losses in passive waveguides warrants further system-level investigation.

Further reduction of the excess optical loss remains important for power-constrained transceivers. A shorter phase shifter can directly reduce propagation loss, with increased bandwidth as an additional benefit; simulations in Fig. 2(h) predict an EO bandwidth exceeding 120 GHz for a 1 mm MZM. At ECOC 2026, a team from IMEC and Ghent University demonstrated 212.5 GBaud PAM4 modulation using a 1 mm silicon MZM with a Vπ·L of 1.8 V·cm [26]. This result further supports the potential of shortening our phase shifter to reduce optical loss while retaining high-speed operation. At the waveguide level, future work could explore more strongly confining rib/ridge-waveguide geometries to reduce optical-mode overlap with the conductive access regions and thereby mitigate the associated optical-loss penalty [27]. In parallel, CMOS-compatible process modules, such as localized nickel silicidation [28] and localized dopant activation [29], could be investigated in the optically less sensitive access regions to further reduce electrical resistance while preserving the active-junction geometry. Together, these approaches provide additional degrees of freedom for jointly optimizing optical loss and electrical bandwidth. Finally, as the measured EO bandwidth exceeds 100 GHz, driver bandwidth and packaging parasitics may become increasingly important system-level constraints. The available setup did not permit 400 Gbit/s PAM4 testing, so the device's system-level data-rate ceiling remains unassessed. Future efforts should therefore prioritize driver–MZM co-design, including differential drive [20,26] with advanced high-speed drivers to increase the delivered voltage swing. In parallel, reducing optical-link loss could lessen the need for PDFA amplification and the associated amplified spontaneous-emission (ASE) noise, while a higher-bandwidth photodiode [30] could further improve end-to-end performance. Together, these measures could translate the modulator's measured bandwidth into higher achievable transmitter data rates.

In summary, using a spatially selective design strategy, we demonstrate a 2 mm stepped-slab silicon MZM on a 300 mm silicon photonics platform, achieving a directly measured 3 dB EO bandwidth of 104 GHz, a modulation efficiency of $V\pi \cdot L$ = 1.5 V·cm, and an open eye diagram at 170 GBaud PAM8 (510 Gbit/s). Wafer-level measurements further yield mean EO bandwidths of 92.8 GHz at −3 V and 100.8 GHz at −5 V, together with a mean $V\pi \cdot L$ of 1.5 V·cm, demonstrating wafer-level uniformity and supporting the suitability of the approach for scalable implementation on a 300 mm silicon photonics platform. This process-compatible approach offers a practical route towards 400 Gbit/s-per-lane silicon photonic transceivers. More broadly, the same spatially selective design strategy can be extended to other carrier-dispersion-based modulators. For silicon-based MZMs, employing shorter, sub-millimeter modulators combined with photonic crystals and frequency-domain equalization could extend the usable bandwidth of all-silicon modulators towards 200 GHz.

## 6. CONCLUSION

## 7. Methods

### Simulation

Large-signal modulation was characterized using the setup shown in Fig. 4(a). A Keysight M8199B pointing to the color features, numbering, etc. The doping profile of the PN junction in MZMs was simulated by Technology Computer Aided Design (TCAD). The series resistance and junction capacitance were obtained with TCAD and imported into HFSS. Combined with a T-shaped traveling electrode model, a Driven Terminal analysis was performed to obtain the S-parameters. Separately, the same doping profile was used to extract the carrier distribution, which was then imported into Lumerical CHARGE and MODE for finite-difference eigenmode (FDE) analysis, yielding the optical mode profile, propagation loss, and modulation efficiency.

## Fabrication

The MZMs were fabricated on a commercial standard CMOS silicon-photonic platform, with 220 nm Si and 2 μm buried oxide layers. The foundry platform offers three etch depths using three mask layers: a 70 nm etch and a 150 nm etch for grating couplers and rib waveguides, and a 220 nm full etch for strip waveguides. The stepped-slab structure was formed by combining the 70 nm and 150 nm etch steps. Additionally, a triple-graded symmetric doping scheme was employed to balance optical loss and electrical performance. The concentrations of the P-type and N-type doping in the waveguide center, which determine the modulation efficiency and dominate the optical loss, are $6.7 \times 10^{17}$ $cm^{-3}$. The intermediate P+/N+ regions and the outermost P++/N++ regions have concentrations of $2 \times 10^{19}$ $cm^{-3}$ and $1 \times 10^{20}$ $cm^{-3}$, respectively. Metal contacts were positioned above the 220 nm-thick Si waveguide. A three-metal stack (Metal1/Metal2/Metal3) and vias (CT/V1/V2) were used for electrical connections, with aluminum (Al) pad arrays arranged at a 100 μm pitch.

## Experimental Details

Eye diagrams of the MZMs were measured with a 256 GSa/s arbitrary waveform generator (AWG, Keysight M8199B) and captured by a digital communication analyzer (DCA, Keysight DCA-X N1000A) through a 90 GHz optical module (Keysight N1032A). The AWG and DCA were triggered by a common clock. For eye-diagram acquisition, waveforms were averaged to suppress PDFA amplified spontaneous emission (ASE) noise, whereas unaveraged waveforms were used for BER calculation. Owing to the limited setup bandwidth, severe high-frequency attenuation markedly reduced the eye opening, making 380 Gbaud PAM4 eye diagrams barely discernible and causing a rapid BER increase (Fig. 4(c)). An external high-frequency SiGe driver [20] is expected to increase the signal swing and yield clearer eye diagrams for 400 Gbps PAM4.

**Acknowledgment.** This work is supported by Zhangjiang laboratory.

**Disclosures.** The authors declare no conflicts of interests.

**Data availability.** Data underlying the results presented in this paper areavailable from the corresponding authors upon reasonable request.

# Supplementary Information – Beyond 100 GHz Silicon Mach-Zehnder Modulators Enabled by Spatially Selective Optimization

**Xingyu Liu,** [1,†] **Fengxin Yu,** [2,†] **Xiao Hu,**[2,*] **Zixuan Cai,**[1] **Haibin Zhao,**[1] **Haiwen Cai,**[2,3] **and Wei Chu**[2,4]
[1]*College of Future Information Technology, Fudan University, Shanghai 200433, China*
[2]*Zhangjiang Laboratory, Shanghai 201210, China*
[3]*caihw@zjlab.ac.cn*
[4]*chuwei@zjlab.ac.cn*
[†]*These authors contributed equally.*
**hxhustoe@163.com*

## S1. Small-Signal RC Model of the PN Junction

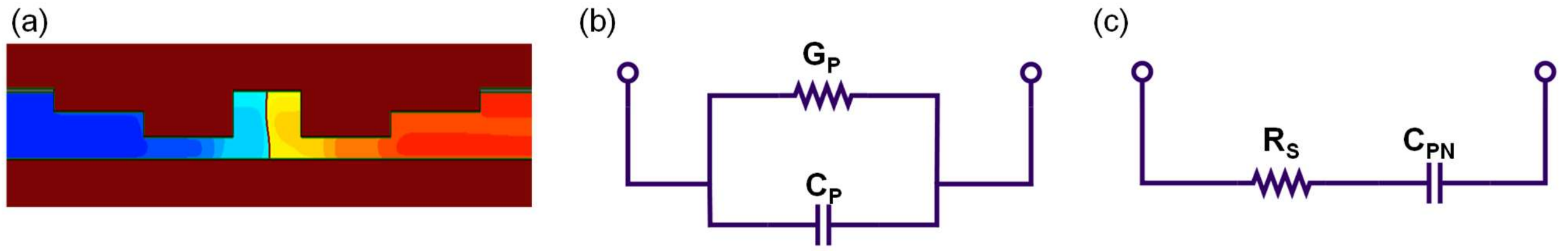

Fig. S1. Small-signal RC model of the PN junction. (a) Cross-section view of PN junction. Equivalent circuits of a reverse-biased PN junction. (b) Parallel RC circuit model. (c) Series RC circuit model.

TCAD process and device simulations were used to obtain a process-representative doping profile and the corresponding carrier distribution of the PN junction (Fig. S1(a)), from which its electrical characteristics were extracted under different reverse-bias voltages and doping concentrations. In TCAD simulations, the semiconductor equations are linearized around the DC operating point, and the terminal-current response to a small AC voltage perturbation is calculated. For the two-terminal junction, the small-signal admittance is expressed as

$$Y = \frac{\delta I}{\delta V} = G_p + j\omega C_p \tag{S1}$$

where $G_p$ and $C_p$ are the equivalent parallel conductance and capacitance, respectively, and ω is the angular frequency. As illustrated in Fig. S1(b), $G_p$ represents the in-phase small-signal current response, whereas $C_p$ primarily reflects charge modulation in the depletion region under reverse bias. The corresponding series impedance is obtained from Z=1/Y, as illustrated in Fig. S1(c). The equivalent series resistance $R_S$ and capacitance $C_{PN}$ are therefore given by

$$Z = \frac{1}{Y} = R_S - \frac{j}{\omega C_{PN}} \tag{S2}$$

$$R_S = \frac{G_p}{G_p^2 + \left(\omega C_p\right)^2} = \frac{R_p}{1 + Q_p^2} \tag{S3}$$

$$C_{PN} = \frac{G_p^2 + (\omega C_P)^2}{\omega^2 C_P} = C_P\left(1 + \frac{1}{Q_p^2}\right) \tag{S4}$$

where $R_p = 1/G_p$, $Q_p = \omega R_p C_p$.

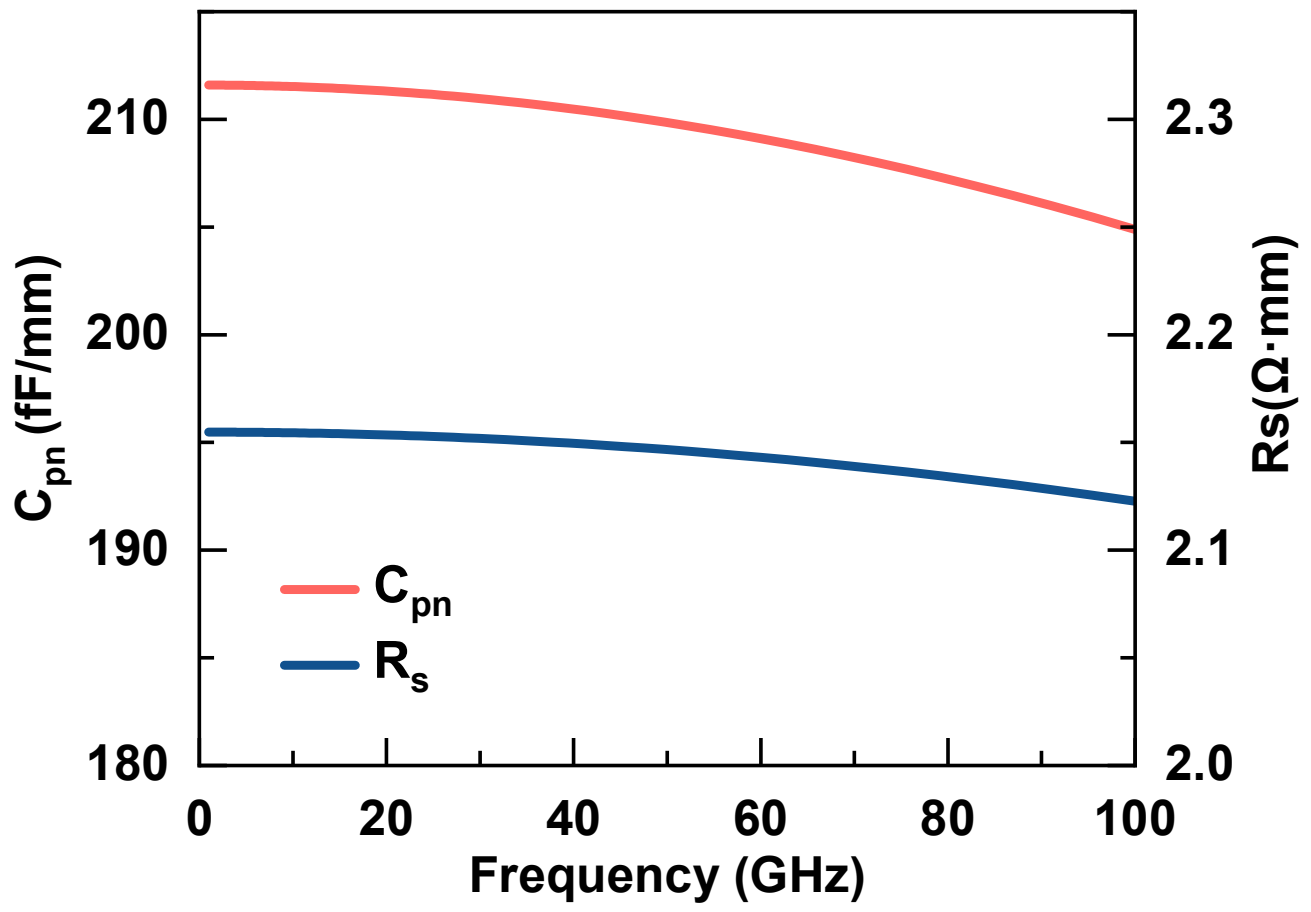

Fig. S2. Small-signal series resistance and capacitance versus frequency

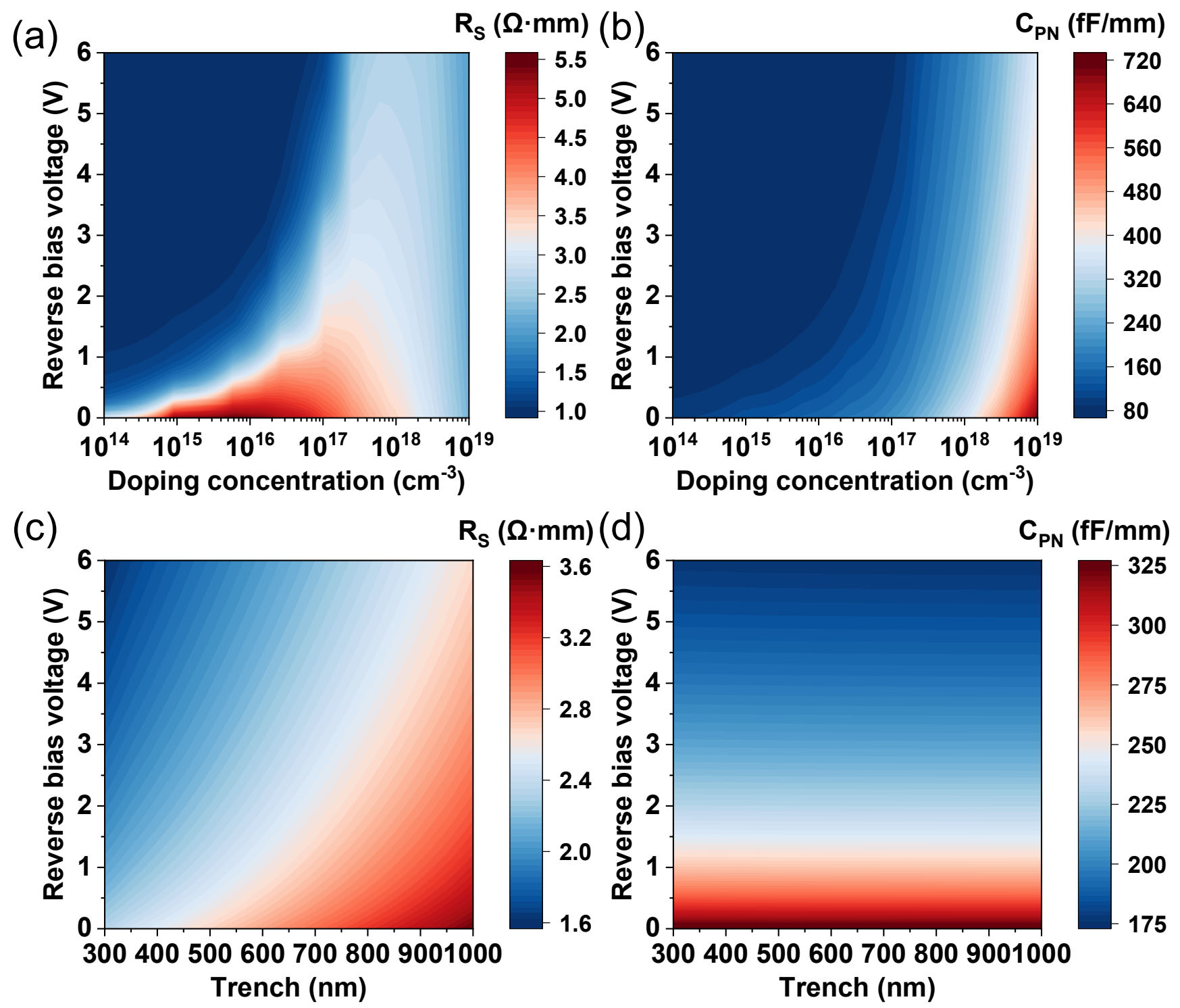

Fig. 3. TCAD simulation results. (a), (b) Rs, Cpn vs. doping concentration and reverse bias. (c), (d) Rs, Cpn vs. trench distance and reverse bias.

Fig. S2 shows the extracted $R_S$ and $C_{PN}$ as functions of microwave frequency from 1 to 100 GHz under −3 V bias. Both $R_S$ and $C_{PN}$ decrease slightly with increasing frequency, with variations of 1.5% and 3.2%, respectively, exhibiting only weak frequency dependence over the examined high-frequency range. Therefore, all parameters are extracted at a fixed frequency of 1 GHz for subsequent analysis. Figs. S3 (a),(b) show $R_S$ and $C_{PN}$ obtained by sweeping the junction doping concentration, whereas Figs. S3 (c),(d) show the corresponding results for the stepped-slab trench geometry. The extracted $C_{PN}$ increases markedly with doping concentration but remains nearly unchanged with trench variation. This indicates that the capacitance is primarily governed by the charge response of the central junction depletion region, which is only weakly affected by changes to the stepped-slab geometry away from the junction. For both parameter sweeps, the extracted $C_{PN}$ and $R_S$ decrease with increasing reverse-bias voltage. The decrease in $C_{PN}$ is consistent with depletion-region widening under reverse bias. The reduction in Rs, however, depends on the combined bias dependence of $G_p$ and $C_P$ in Eq. (S3), rather than on the capacitance reduction.

## S2. Electro-Optic Response Model

Following the segmented traveling-wave model reported in Ref. S1, the traveling-wave Mach–Zehnder modulator (MZM) is divided into N uniform sections, each incorporating a PN-junction-loaded segment of length $l_0$. Additional unloaded electrode sections of lengths $l_1$ and $l_2$ are included at the input and output, respectively. The ABCD transfer matrices of all sections are cascaded to obtain the overall response, as expressed in Eq. (S5).

$$M(f) = \left| \frac{2}{NV_S} \sum_{n=1}^{N} V_n e^{j\omega[l_1+(n-1)l_0]/v_0} \frac{1}{1-\omega^2 L_a C_a + j\omega R_a C_a} \right|^2 \tag{S5}$$

Here, $V_S$ is the input signal voltage, $V_n$ is the microwave voltage at the $n_{th}$ loaded section, and the exponential term $e^{j\omega[l_1+(n-1)l_0]/v_0}$ represents the microwave propagation phase factor. $L_a$, $R_a$, and $C_a$ denote the equivalent series inductance, resistance, and depletion capacitance of each lumped PN-junction section, respectively.

In this model, the effective modulation voltage is the voltage across the depletion capacitance $C_a$. Neglecting the series inductance $L_a$, the effective voltage is given by Eq. (S6):

$$V_{eff} = \frac{V_n}{1 + j\omega R_a C_a} \tag{S6}$$

As the angular frequency increases, the magnitude of the denominator increases, reducing the voltage across $C_a$ and thereby suppressing the high-frequency EO response. Reducing $R_a$ mitigates resistive dissipation in the PN-junction load and reduces the high-frequency voltage drop across the series resistance, preserving a larger effective modulation voltage and improving the overall EO response.

## S3. Peaking of the EO Response Caused by Termination Mismatch

Impedance mismatch at the end of a traveling-wave electrode causes partial microwave reflection, characterized by the load reflection coefficient:

$$\Gamma_L = \frac{R_L - Z_0}{R_L + Z_0} \tag{S7}$$

where $R_L$ and $Z_0$ are the termination resistance and characteristic impedance, respectively. When $R_L < Z_0$, $\Gamma_L < 0$, and the reflected voltage wave undergoes a $180°$ phase inversion. Assuming a matched source and neglecting multiple reflections, the microwave voltage along an electrode of length L is

$$V(z,\omega) = V^+ \left[ e^{-\gamma_m z} + \Gamma_L e^{-\gamma_m (2L-z)} \right] \tag{S8}$$

where $V^+$ is the incident voltage amplitude and $\gamma_m = \alpha_m + j\beta_m$ is the microwave propagation constant.

Following the analytical treatment of traveling-wave MZMs in Ref. S2, the EO response can be decomposed into contributions from co-propagating and counter-propagating microwave waves. For a matched source, a dispersionless transmission line, and negligible microwave attenuation, the normalized EO amplitude response can be expressed as

$$|H_{norm}(f)| = \frac{1}{|1+\Gamma_L|}\left|\mathrm{sinc}\left[\frac{\pi f L}{c}\left(n_{og}-n_e\right)\right] + \Gamma_L e^{-j2\pi f \; eL/c}\,\mathrm{sinc}\left[\frac{\pi f L}{c}\left(n_{og}+n_e\right)\right]\right| \quad \text{(S9)}$$

where $\mathrm{sinc(x)} = \sin(\mathrm{x})/x$, $n_{og}$ is the optical group index, and $n_e$ is the microwave phase index. The first term describes the co-propagating contribution, whereas the second describes the counter-propagating contribution induced by reflection. Because the two waves propagate in opposite directions, the reflected contribution experiences a larger optical–microwave phase mismatch and therefore rolls off more rapidly with frequency.

At low frequencies, both $\mathrm{sinc}$ terms approach unity, and the EO amplitude relative to the matched-termination response is reduced to $1+\Gamma_L$. The corresponding low-frequency gain change is

$$\Delta H_{EO}(0) = 20\log_{10}|1+\Gamma_L| = 20\log_{10}\left|\frac{2R_L}{R_L+Z_0}\right| \quad \text{(S10)}$$

As the frequency increases, the counter-propagating contribution diminishes, allowing the EO response to recover toward its nominal level. When the optical group index approximately matches the microwave phase index, the characteristic recovery frequency is

$$f_{rec} \approx \frac{c}{2\left(n_{og}+n_e\right)L} \quad \text{(S11)}$$

This relation predicts an inverse dependence on electrode length. For $n_{og} \approx n_e \approx 4$, the characteristic frequencies are approximately 9.4 GHz and 37.5 GHz for electrode lengths of 2 mm and 0.5 mm, respectively. These values are close to the simulated response peaks of approximately 11 GHz and 45 GHz, respectively. The agreement supports termination-induced reflection as a contributing mechanism for the observed peaking. The exact peak position and amplitude, however, also depend on the reflection coefficient, microwave attenuation, optical–microwave velocity mismatch, and source-end reflections.

Fig. S4 compares the normalized EO responses of MZMs with different lengths under matched and mismatched termination conditions at a reverse bias of −3 V. As the MZM length decreases, the peaking feature shifts toward higher frequencies. The corresponding apparent 3 dB bandwidth enhancements due to peaking are 12, 14, 16, 12, and >20 GHz for MZM lengths of 2, 1.5, 1, 0.5, and 0.25 mm, respectively. Combined with the reduced microwave propagation loss in shorter electrodes, these results suggest a potential pathway toward 200 GHz bandwidth in silicon-based MZMs.

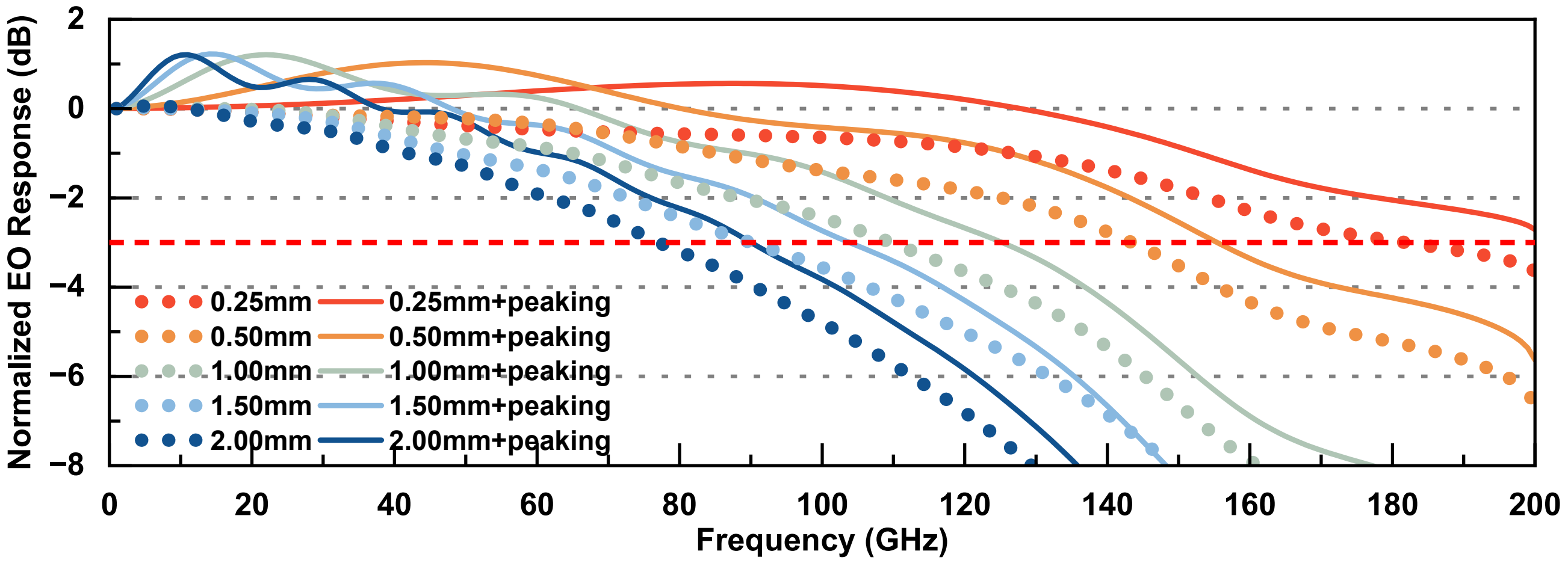

Fig. S4. EO response of MZMs with various lengths under termination matching/mismatch.

## S4. Comparison between Simulated and Measured Device Performance

The fabricated stepped-slab MZM was designed with a trench-to-waveguide-edge distance of 500 nm and a central active-region doping concentration of $6.7\times10^{17}cm^{-3}$. Fig. S5(a) compares the simulated propagation loss of the stepped-slab and conventional waveguides with the same doping concentration, yielding 18.5 and 16.4 dB $cm^{-1}$, (3.7 and 3.24 dB for a 2 mm phase shifter), respectively. The complete MZM exhibits a measured on-chip insertion loss of 4.5 dB, which is 0.8 dB higher than the simulated stepped-slab contribution of 3.7 dB over the same 2 mm length. The remaining difference may arise from excess loss in the MMI splitters and combiners, access and transition waveguides, as well as fabrication-induced deviations in the waveguide geometry and doping profile [3,4]. These contributions cannot be individually quantified from the present measurements.

Fig. S5(b) shows the simulated modulation efficiency as a function of reverse-bias voltage. The $V_\pi \cdot L$ product increases with increasing reverse-bias magnitude, indicating a reduction in modulation efficiency. At −3 V, the simulated $V_\pi \cdot L$ is 1.56 V·cm, in close agreement with the measured value of 1.50 V·cm, with a deviation of approximately 4%.

Fig. S5(c) compares the simulated and measured EO frequency responses. As the reverse bias varies from −1 to −5 V, the simulated 3-dB EO bandwidths are 67, 80, 90, 98, and 106 GHz, respectively, compared with measured values of 63, 81, 94, 98, and 104 GHz. The absolute difference does not exceed 4 GHz at any bias, with a mean absolute difference of 2.2 GHz. The close agreement in both bandwidth values and their bias dependence supports the validity of the combined electrical and EO models.

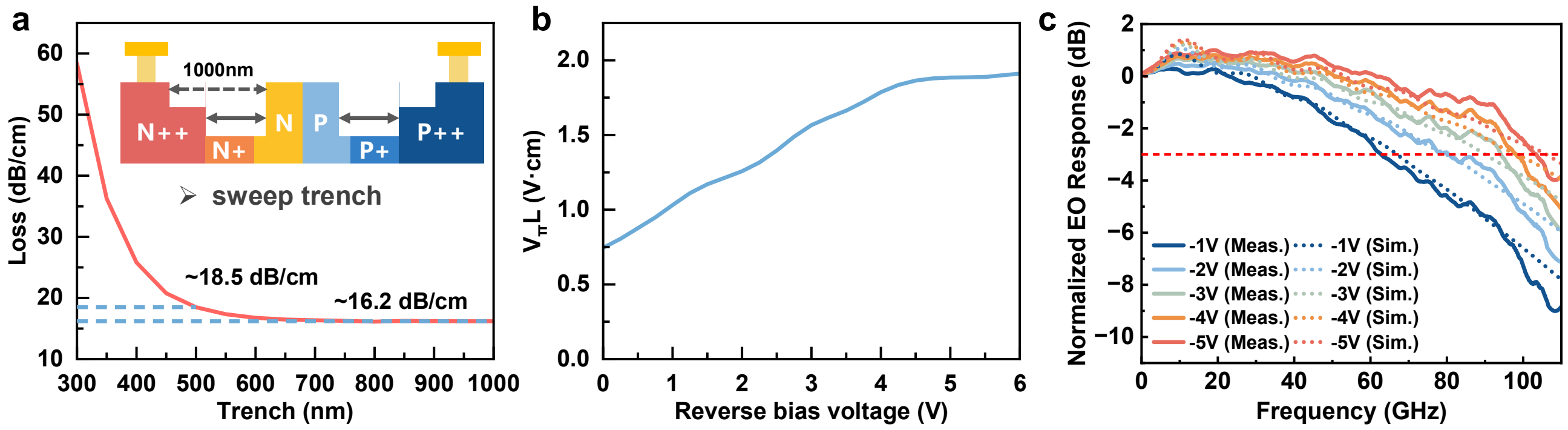

Fig. S5. Comparison of simulated and measured results. (a) Simulated optical loss versus trench distance. (b) Simulated modulation efficiency versus reverse-bias voltage. (c) EO responses at various bias voltages: simulated vs. measured.

## Supplementary References